\documentclass[a4paper,twocolumn,%tightenlines,
english,aps,pre,floatfix,showpacs]{revtex4}
\usepackage[T1]{fontenc}
\usepackage[latin1]{inputenc}
\usepackage{amsmath}
\usepackage{babel}
\usepackage{graphics}
\usepackage{amssymb}

\begin{document}
 
\title{Roughness of time series in a critical interface model
}

\author{S.L.A. \surname{de Queiroz}}

\email{sldq@if.ufrj.br}

\affiliation{Instituto de F\'\i sica, Universidade Federal do
Rio de Janeiro, Caixa Postal 68528, 21941-972
Rio de Janeiro RJ, Brazil}

\date{\today}

\begin{abstract}
We study roughness probability distribution functions (PDFs) of the time signal for 
a critical  interface model, which is known to provide a good description of 
Barkhausen noise in soft ferromagnets. Starting with time ``windows'' of data 
collection much larger than the system's internal ``loading time'' 
(related to demagnetization effects), we show that the initial Gaussian shape of 
the PDF evolves  into a double-peaked structure as window width decreases. We 
%%%%%%%%%%%%%%%% 2nd ref's point # 7 %%%%%%%%%%%%%%%%%%%%%%%%%%%%%%%%%%%
advance a plausible  physical explanation for such structure, which is broadly compatible 
with the observed numerical data. 
%%%%%%%%%%%%%%%%%%%%%%%%%%%%%%%%%%%%%%%%%%%%%%%%%%%%
Connections to experiment are suggested.
\end{abstract}
\maketitle
%\tightenlines
\section{INTRODUCTION}
\label{intro}

The probability distribution functions (PDFs) of critical fluctuations in assorted
systems have been the subject of much recent interest, mainly stemming from the
realization that they exhibit a remarkable degree of
universality~\cite{bhp98,bhp00,adgr01,adgr02}.

The roughness $w_2$ of a fluctuating interface with $N$ elements is  
the position-averaged square  width  of the interface height above an arbitrary 
reference level~\cite{adgr02,rkdvw03}: 
\begin{equation}
w_2  =N^{-1}\,\sum_{i=1}^N\left(h_i-\overline{h} \right)^2\ ,
\label{eq:rough1}
\end{equation}
where $\overline{h}$ is the average interface height. 
The finite-size scaling of the first moment of the roughness PDF gives the 
roughness exponent $\zeta$~\cite{bar95}:
\begin{equation}
\langle w_2 (L)\rangle \sim L^{2\zeta}\ \ ,
\label{eq:zeta}
\end{equation}
where angular brackets stand for averages over the ensemble of allowed interface 
configurations, and $L$ is some finite linear dimension characterizing the system
in study. The width 
%%%%%%%%%%%%%%%%%%%%%%%%%%%%% 2nd ref's point # 1 %%%%%%%%%%%%%%%%%%%%%%%%%%%%%
PDF $P(w_2)$ 
%%%%%%%%%%%%%%%%%%%%%%%%%%%%%%%%%%%%%%%%%
for correlated systems at 
criticality may be put into a scaling form~\cite{forwz94,adgr01,adgr02,rkdvw03}, 
\begin{equation}
\Phi (z) = \langle w_2\rangle\,P(w_2)\ ,\quad z \equiv w_2 /\langle 
w_2 \rangle\ ,
\label{eq:pvsphi}
\end{equation}
%%%%%%%%%%%%%%%%%%%%%%%%%%%%% 2nd ref's point # 1 (cont)%%%%%%%%%%%%%%%%%%%%%%%%%%%%%
i.e., the scaling function $\Phi (z)$ is expected to depend only on the scaled width 
$  w_2 /\langle w_2 \rangle$. In other words, the size dependence must appear
%%%%%%%%%%%%%%%%%%%%%%%%%%%%%%%%%%%%%%%%%%%%%%%%%%%%%%%%%%%%%
exclusively through the average width $\langle w_2\rangle$. 
Comparison of experimental or simulational data to specific analytical forms, 
whose suitability to the description of the case at hand has been anticipated by 
physical arguments, usually results in good agreement. 
Thus, the PDF of voltage  fluctuations in 
semiconductor films was fitted very well by that of perfect Gaussian $1/f$ 
noise~\cite{adgr01}; simulational data for the single-step model of 
deposition-evaporation, by the PDF of a random-walk process~\cite{forwz94} 
(the latter corresponds to perfect Gaussian $1/f^2$ noise, or {\it Wiener 
process}~\cite{adgr02}). Further progress was made possible via the 
analytical  evaluation of  roughness PDFs for generalized  Gaussian noise with 
independent Fourier modes ( i.e. $1/f^\alpha$ noise with general, continuously-varying 
$\alpha$)~\cite{adgr02}~. Consideration of the scaling properties of height-height 
correlation functions and their Fourier transforms implies~\cite{rkdvw03} that
\begin{equation}
\alpha= d + 2\zeta\qquad \ ,
\label{eq:alphazeta}
\end{equation}
where $d$ is the interface dimensionality and $\zeta$ is defined in    
Eq.~(\ref{eq:zeta})~. Eq.~(\ref{eq:alphazeta}) is valid provided that
$\zeta >0$~\cite{adgr02}, i.e. $\alpha >1$ in the present case where the 
``interface" is a time series (see below).	 

In previous work~\cite{rdist} we applied the ideas outlined above, to
investigate the interface roughness PDFs of a single-interface model which has
been used in the description of Barkhausen  ``noise'' (BN)~\cite{umm95,{us,us2,bark3}}.
This is an intermittent phenomenon which reflects the
dynamics of domain-wall motion in the central part of the hysteresis cycle
in ferromagnetic materials (see Ref.~\onlinecite{dz05} for an up-to-date 
review). By ramping an externally applied magnetic field, one causes
sudden turnings (avalanches) of groups of spins. The consequent changes in
magnetic flux induce a time-dependent electromotive force $V(t)$ on a coil
wrapped around the sample. Analysis of $V(t)$, assisted by suitable
theoretical modeling, provides insight into both the domain structure itself and 
its dynamical behavior. 
It has been proposed that BN is an illustration of ``self-organized
criticality''~\cite{bw90,cm91,obw94,umm95}, in the sense that a broad 
distribution of scales (i.e. avalanche sizes) is found within a wide range 
of variation of the external parameter, namely the applied magnetic field,
without any fine-tuning. The interface model studied here~\cite{umm95}
incorporates a self-regulating mechanism, in the form of a demagnetization 
factor. 

We have shown~\cite{rdist} that the demagnetizing term is irrelevant as regards
interface roughness distributions, with the conclusion that in this respect the
behavior of self-regulated systems is in the same universality class as that of the
quenched Edwards-Wilkinson model~\cite{les93,ma95,mbls98,rhk03}, at criticality
(i.e. at the interface depinning transition).

However, when one considers the time series of intermittent events which
characterizes BN, it is known that the demagnetizing term is responsible for
the introduction of short-time negative correlations in the model (such
correlations are observed in experiments as well)~\cite{umm95}.
The question then arises of whether a corresponding signature of self-regulation
will be present when the roughness distribution of the time sequence of BN events 
is  examined. Since the traditional data acquisition method in the study of BN is 
exactly via the time series of induced voltages, an 
investigation  along these lines may establish useful connections between 
observational data and the basic physical mechanisms underlying BN.      

\section{Model ingredients and dynamics}
\label{sec:2}
 
We use the single-interface model introduced in
Ref.~\onlinecite{umm95} for the description of BN. In line with
experimental procedure, the external field $H$
acting on the sample is assumed to increase linearly in time, therefore
its value is a measure of ``time".  We consider the
adiabatic limit of a very slow driving rate, thus avalanches are
considered to be instantaneous (occurring at a fixed value of the external
field). In this simplified version, a plot of $V(t)$ against $t$ 
consists of a series of spikes of varying sizes, placed at non-uniform 
intervals. Generalizations for a finite driving rate may be 
devised~\cite{us2,tad99,wd03}, 
but will not concern us here.

Simulations are performed on an $L_x \times L_y \times \infty$  geometry,
with the interface motion set along the infinite direction. Since we are
interested in fluctuations of the Barkhausen signal in {\it time}, we keep 
geometric aspects at the simplest level, i.e. $L_y=1$ (system 
dimensionality $d=2$, interface dimensionality $d^\prime = 1$). Periodic boundary 
conditions are imposed at $x=0,\,L$~. 

%%%%%%%%%%%%%%%%%%% 2nd ref's point # 2 %%%%%%%%%%%%%%%%%%%%%%%%%%%%
The interface ($180$-degree domain wall separating spins parallel to the external field
from those antiparallel to it) is composed by $L$ discrete elements whose $x$ coordinates
are $x_i=i$, $i=1, \dots, L$, and whose (variable) heights above an arbitrary reference 
level are $h_i$. The simulation starts with a flat wall: $h_i=0$ for all $i$.
%%%%%%%%%%%%%%%%%%%%%%%%%%%%%%%%%%%%%%%%%%%%%%%%%%%%%%%%%%%%%%%%%%%%%%%%%
  
Each element $i$ of the interface experiences a force given by:
\begin{equation}
f_i=u(x_i,h_i)+{k}\,\left[h_{i+1} + h_{i-1}- 2\,h_i\right]+H_e~,
\label{force}
\end{equation}
where
\begin{equation}
H_e =H -\eta M~.
\label{He}
\end{equation}
The first term on the RHS of Eq.~(\ref{force}) represents quenched disorder, and
is drawn from a Gaussian distribution of zero mean and width $R$;
the intensity of surface tension is set by $k$, and the effective field $H_e$
is the sum of a time-varying, spatially uniform, external field $H$ and a 
demagnetizing field which is taken to be proportional to
$M=(1/L)\sum^{L}_{i=1} h_i$, the magnetization  (per site) of the previously 
flipped spins for a lattice of transverse width $L$.
Here we use $R=5.0$, $k=1$, $\eta=0.005$, values for which fairly broad
distributions of avalanche sizes are obtained~\cite{rdist,us,us2,bark3}.

The dynamics goes as follows. For fixed $H$, starting from zero, the sites are 
examined sequentially; at those for which $f_i > 0$, $h_i$ is increased by one 
unit, with $M$ being  updated accordingly; the corresponding new value of $u$ is 
drawn. The whole  interface is swept as many times as necessary, until only sites 
with $f_i < 0$ are left, which marks the end of an avalanche. The external field is 
then increased until $f_i=0$ for at least one site. This is the threshold
of a new avalanche, which is triggered by the update of the site(s) with $f_i=0$, 
and so on.   

The effect of the demagnetizing term on the effective field $H_e$ is that at
first it rises linearly with the applied field $H$, and then, upon further increase
in $H$, saturates (apart from small fluctuations)  at a value rather close to the
critical external field for the corresponding model {\em without}
demagnetization~\cite{umm95,us}.

\section{Time series: correlations and roughness}
\label{sec:3}

As explained above, owing to the assumed linear increase of applied field  
with time (in analogy with experimental setups), we shall express time in
units of $H$ as given in Eqs.~(\ref{force}) and~(\ref{He}).
We have generated time series of BN, with  ${\cal O}\,(10^4 - 10^5)$
events. Steady state, i.e., the stabilization of $H_e$ of Eq.~(\ref{He})
against external field $H$, occurs after some 200 events, for the range of
parameters used here. Though we used only steady-state data, it was noted
that inclusion of those from the transient does not appreciably distort
any of the quantities studied.

In experiment, the integrated signal $\int_{\Delta t} V(t)\,dt$ is
proportional to the magnetization change (number of upturned spins) during
the interval $\Delta t$. In the adiabatic approximation used here, a
box-like shape is implicitly assumed for each avalanche
%%%%%%%%%%%%%%%%%%%%%% 2nd ref's point # 3 %%%%%%%%%%%%%%%%%%%%%%%%%%%%%%%%%%%%%%
(i.e. details of the internal structure of each peak, as it develops in time, are 
ignored, on acount of its duration being very short),
%%%%%%%%%%%%%%%%%%%%%%%%%%%%%%%%%%%%%%%%%%%%%%%%%%%%%%%%%%%%
thus the instantaneous signal intensity (spike height) is proportional to the
corresponding avalanche size.

We sample the fluctuations of the signal  along successive ``windows" of
equal time duration $W$, each containing many spikes. Each window is divided
into equally-spaced bins of size $\delta$; the signal intensity associated to each  bin
is the sum of the sizes of all avalanches which occurred  within that bin.
%%%%%%%%%%%%%% 1st ref's suggestion # 1 %%%%%%%%%%%%%%%%%%%%%%%%%%%%%%%%%%%%%%%%%
The roughness $w_2$ of the signal on a given window starting, say, at $t=0$, is given 
by
\begin{equation}
w_2 = \frac{1}{W/\delta}\sum_{i=1}^{W/\delta} \left(V_i-\overline{V} \right)^2\ ,
\ \ V_i =\sum_{t \in [(i-1)\delta, i\,\delta]}V(t)\ ,
\label{eq:trough}
\end{equation}
where $\overline{V}$ is an average of $V(t)$ over the whole window span $W$.
%%%%%%%%%%%%%%%%%%%%%%%%%%%%%%%%%%%%%%%%%%%%%%%%%%%%%%%%%%%%%%%%%%%%%%%%%%%%%%%

As the signal is intermittent, there are significant periods (waiting
times, henceforth referred to as WT) of no activity at all. Such quiet
intervals must be properly accounted for in the statistics of
fluctuations, hence care must be taken when setting up the bin
size~$\delta$. 

\begin{figure}
{\centering \resizebox*{3.4in}{!}
{\includegraphics*{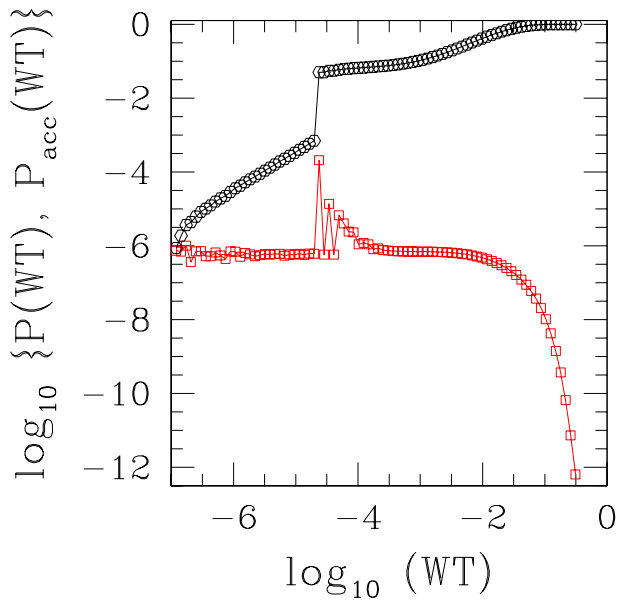}} \par}
\caption{(Color online) Double-logarithmic plot of probability distribution, $P({\rm 
WT})$, of waiting times (lower curve), and accumulated distribution, $P_{\rm acc}({\rm 
WT}) \equiv \int_0^{\rm WT} P(t)\,dt$ (upper curve). $L=400$.
}
\label{fig:wtdist}
\end{figure}

We have examined WT distributions, for varying lattice widths $L =
200,\ 400,\ 800$.
In Fig.~\ref{fig:wtdist}  (lower curve) we display a double-logarithmic plot of the 
probability of occurrence of assorted WTs for $L=400$, against 
%%%%%%%%%%%%%%%%%%%%%%%%%%%%%%%%%% 2nd ref's point # 4 %%%%%%%%%%%%%%%%
${\rm WT}$, 
%%%%%%%%%%%%%%%%%%%%%%%%%%%%%%%%%%%%%%%%%%%%%%%%%%%%%%%%%%%%%%%%%%%%%%%
sampled over $8 \times 10^6$ events. The distribution is generally rather flat,
apart from (i) a sharp cutoff at the high end (related to the finite
cutoff in the avalanche size probability distribution, see the discussion
of loading times below), and (ii) a number of
peaks concentrated in a somewhat narrow region corresponding to $10^{-5}
\lesssim {\rm WT} \lesssim 10^{-4}$. The latter are associated to very
frequent and small, spatially localized  (i.e. non-critical) events
involving typically $N = 1-10$ sites~\cite{us2}. 
This is easy to see by recalling
from Eqs.~(\ref{force}) and~(\ref{He}) that, since the demagnetization
term keeps $H_e$ approximately constant, a small avalanche with $N$
spins overturned decreases the internal field by $\eta\,N/L$, thus
requiring approximately the same increase in external field in order to
bring the system back to criticality. We have checked that the peaks
move consistently with this argument, i.e. their horizontal position
is shifted leftward by a factor of $\log_{10} 2$ for each doubling of $L$. 

Upon consideration of integrated WT distributions (upper curve in 
Fig.~\ref{fig:wtdist}), we decided to set the bin size $\delta=10^{-5}$ (for 
$L=400$, the system size for which most of our calculations are done, see below). 
With such  a choice, WTs shorter than $\delta$ occur with less than $1\%$  
frequency. This ensures both that inactive periods are not wrongly obscured by
bursts of activity, and that consecutive avalanches are rather unlikely to
be lumped together.   

%%%%%%%%%%%%%%%%%%%%%% 1st ref's Comment %%%%%%%%%%%%%%%%%%%%%%%%%%%%%%%%%%%%%%%%%%%
At this point, a comment must be made on the connection of the above results with 
previous investigations of WT distributions in BN. In Ref.~\onlinecite{dbm95} it was
predicted, from a fractal analysis of the  ABBM~\cite{abbm2} model which 
describes domain-wall motion via a Langevin equation, that $P({\rm WT}) 
\sim ({\rm WT})^{-(2-c)}$, where $c$ is proportional to the external field driving rate.
Experimental data in SiFe samples are consistent with this~\cite{dbm95}.
The present case of adiabatic driving ~~would then correspond to $c \to
0$. However, it is 
crucial in the analysis of Ref.~\onlinecite{dbm95} that the BN pulse durations be
finite (even though they shorten accordingly in the $c \to 0$ limit). Indeed, the result
just quoted relies on considering the properties of complementary sets, both with  
non-zero fractal dimension (namely the time intervals during which there is domain-wall 
motion, versus those of no activity, i.e. WTs). 
The approximation used here, of considering BN pulses as having exactly zero duration, 
destroys the connection of our data with the conceptual framework in which the power-law 
dependence  $P({\rm WT})\sim ({\rm WT})^{-(2-c)}$ was found. Though in this sense the 
flat WT distribution found here is most likely an artifact of the model, the conclusions 
extracted from the distribution with respect to the choice of $\delta$ remain
valid.       
%%%%%%%%%%%%%%%%%%%%%%%%%%%%%%%%%%%%%%%%%%%%%%%%%%%%%%%%%%%%%%%%%%%%%%%%%%%%%%%%

\begin{figure}
{\centering \resizebox*{3.4in}{!}
{\includegraphics*{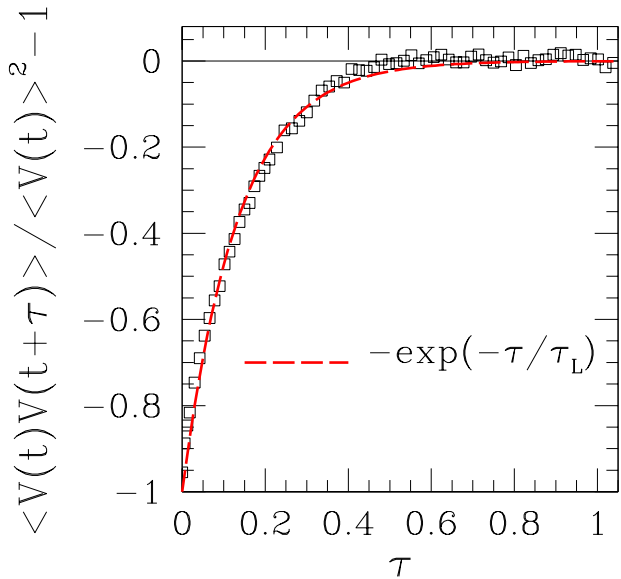}} \par}
\caption{(Color online) Normalized two-time correlations  (averaged over $t$) $\langle
V(t)\,V(t+\tau) \rangle/\langle V(t) \rangle^2 -1$  
for system with $L=400$. Dashed line is
fit of data to single-exponential form, from which $\tau_L=0.14(1)$~. 
}
\label{fig:tcorr}
\end{figure}

We now turn to the choice of window width $W$. Recall that real-space
properties, e.g. interface roughness, of the systems under study
benefit from divergence of the system's natural length scale, as
it self-tunes its behavior to lie close to a second-order
(depinning) transition~\cite{rdist}.
For  such quantities, universality ideas apply, so one expects finite-lattice
effects to be present only as an overall scale factor. e.g. 
$\langle w_2\rangle$ in 
Eq.~(\ref{eq:pvsphi})~\cite{forwz94,adgr01,adgr02,rkdvw03},
However, in the study of time series for the same systems, one must bear in
mind that a finite time scale $\tau_L$ (``loading time'') is introduced
via the demagnetization term~\cite{umm95}. This is illustrated in 
Fig.~\ref{fig:tcorr} (similar  plots, exhibiting both simulational and experimental 
results, can be found in Ref.~\onlinecite{umm95})
where normalized two-time correlations $\langle V(t)\,V(t+\tau) \rangle
/\langle V(t) \rangle^2 -1$ (averaged over $t$) are shown. 
Therefore, different regimes will be found, depending on the value of $x 
\equiv W/\tau_L$. The limit $x \gg 1$ is expected to reproduce
the white noise characteristic of uncorrelated 
fluctuations, for which the roughness distribution is a pure Gaussian. On the other 
hand, non-trivial effects may arise for $x \sim 1$.
 
Before going further, it must be remarked that $\tau_L$ in fact 
decreases for  increasing $L$.
This can be understood by recalling that (i) the
probability distribution for avalanche size $s$ goes roughly as $P(s) \sim
s^{-\tau}\,\exp(-s/s_0)$~\cite{umm95,us,us2,bark3}; (ii) the cutoff $s_0$
scales approximately as $s_0 \sim L^{0.8}$ in the present case
of a one-dimensional interface~\cite{us}.
Thus the maximum waiting time $\tau_M$ will vary as $\tau_M =\eta\,s_0/L
\sim L^{-0.2}$. For $L=400$, we find $s_0 \simeq 4 \times
10^4$~\cite{bark3},
which explains both the sharp drop in the WT distribution at ${\rm WT}
\simeq 0.5$ in Fig.~\ref{fig:wtdist}, and the complete vanishing of
correlations at $\tau \gtrsim 0.5$ in Fig~\ref{fig:tcorr}.
   
In BN studies the connection between lattice-size-dependent 
quantities in simulations, and their experimental counterparts, becomes especially 
clear when one considers the $L$-dependent cutoff in the power-law avalanche size 
distribution, and its relationship to the maximum domain size in 
magnetic samples~\cite{us}. In the present case it should be stressed that finite 
loading times are measured in experiment, under suitable 
conditions~\cite{umm95}. Thus, we assume that the loading times found here are not 
simply a  finite-size artifact of simulations, bound to vanish in the thermodynamic 
limit characteristic of real systems. Instead, although we are not in a 
position to propose quantitative comparisons,  they must correspond to
the experimentally-observed ones . 

\section{results}
\label{sec:results}
%%%%%%%%%%%%%%%%%% 1st ref's suggestion # 1 (cont) #################
By generating many realizations of the roughness $w_2$ defined in 
Eq.~(\ref{eq:trough}) for given values of the physical parameters, we have
obtained the corresponding roughness PDFs.
%%%%%%%%%%%%%%%%%%%%%%%%%%%%%%%%%%%%%%%%%%%%%%%%%%%%%%%%%%%%%%%%%%%%%%%%%%%
The shapes of roughness PDFs found here do not usually
conform to the generalized Gaussian ($1/f^\alpha$) distributions introduced in
Ref.~\onlinecite{adgr02}, although they display certain similarities to the pure
Gaussian limit, which corresponds to $\alpha=1/2$ in the scheme of
Ref.~\onlinecite{adgr02}. We have found it convenient to adhere 
to conventions used in that Reference and related work, namely expressing
the  PDFs in a scaling form, see Eq.~(\ref{eq:pvsphi}). 

We first examine the limit $x \gg 1$. Similarly
to $1/f^\alpha$ PDFs with $\alpha \leq 1$~\cite{adgr02}, our results in this
limit  approach a $\delta$- function shape
when expressed in terms of $z$ of Eq.~(\ref{eq:pvsphi}).
The solution, pointed  out in Refs.~\onlinecite{adgr01,adgr02}, is to 
use scaling by the variance, instead of by the average, i.e. switch to 
the variable 
\begin{equation}
y = \frac{w_2 - \langle w_2 \rangle}{\sqrt{\langle w_2^2 \rangle - \langle w_2 
\rangle^2}}\ .
\label{eq:y}
\end{equation} 
The corresponding scaling function will be denoted by $\Psi (y)$.

\begin{figure}
{\centering \resizebox*{3.4in}{!}
{\includegraphics*{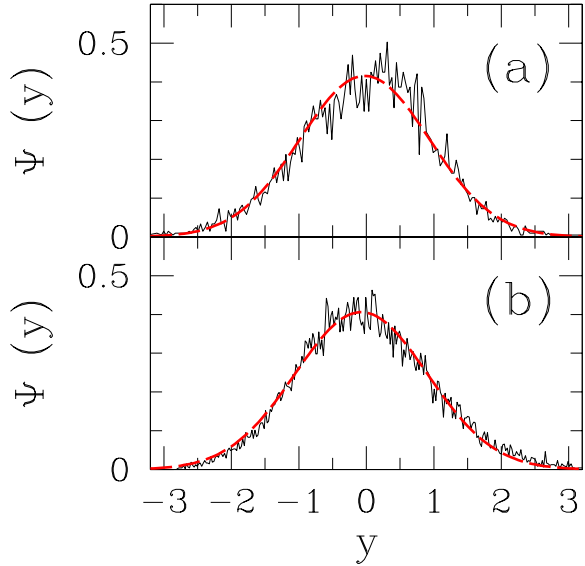}} \par}
\caption{(Color online) Scaled roughness distributions $\Psi (y)$ of time series, for 
$y$ of  Eq.~(\protect{\ref{eq:y}}). $L=400$, window width $W=100$.
(a): demagnetizing factor $\eta=0.005$ ($\tau_L \simeq 0.14$), $6 \times 10^3$ 
samples. Dashed line is 
Gaussian fit to data with mean at $y=0.04(1)$, width $\sigma=0.96(1)$.
(b): demagnetizing factor $\eta=0$ (see text),  $2.1 \times 10^4$
samples. Dashed line is Gaussian fit to data, with mean at $y=-0.07(1)$, width 
$\sigma=0.98(1)$.
}    
\label{fig:ltrt}
\end{figure}

In Fig.~\ref{fig:ltrt} we show results for  window width $W=100$, in terms of
$y$ of Eq.~(\ref{eq:y}). While in (a) the 
demagnetizing factor is $\eta =0.005$ 
(thus $\tau_L \simeq 0.14$ from Fig.~\ref{fig:tcorr}), the data in (b) correspond
to simulations of the same system, with $\eta=0$. As explained in 
Ref.~\onlinecite{rdist}, in this case the system is kept close to criticality
by the following procedure. We first determined the approximate critical
value $H_e^c$ of the internal field $H_e$ of Eq.~(\ref{He}), by starting a
simulation with $\eta \neq 0$ and waiting for $H_e$ to stabilize. At that
point, we set $\eta=0$ and repeatedly swept $H$ in the interval
$(\gamma\,H_e^c, H_e^c )$, $\gamma \lesssim 1$, according to the procedure
delineated in Sec.~\ref{sec:2}. We have used $\gamma=0.9$ for the data displayed in
Fig.~\ref{fig:ltrt}~(b). With $H_e^c \simeq 5.4$ for the disorder and elasticity
parameters used here, data corresponding to a window of ``width'' $W=100$ in this 
case was in fact given by the collation of data from $\sim W/(1-\gamma)\,H_e^c 
=185$ consecutive field sweeps as just described.
%%%%%%%%%%%%%%%% 1st ref's suggestion # 2 part I #############################
Note that, within a given field sweep, many non-critical events are thus sampled (which
would by themselves give rise to a non-universal PDF, see below the discussion for 
narrow windows). However, owing to the central limit theorem, the result 
of the  collation of many independent segments  should yield an overall behavior which 
is essentially Gaussian.  
%%%%%%%%%%%%%%%%%%%%%%%%%%%%%%%%%%%%%%%%%%%%%%%%%%%%%%%%%%%%%%%%%%%%%%%%%%%%%%%%%%   

One can see that in both cases, a single Gaussian centered at $y \simeq 0$
and with variance $\simeq 1$ gives a good fit to data, confirming our expectation
that demagnetization-induced correlations would be essentialy washed away for $W 
\gg \tau_L$. It is worth mentioning, however, that the {\em unscaled} 
variables tell a slightly different story: for the data of Fig.~\ref{fig:ltrt}~(a)
one has $\langle w_2\rangle \pm \sigma = (127 \pm 6)\times 10^3$, while in (b)   
$\langle w_2\rangle \pm \sigma = (6.3 \pm 2.2)\times 10^3$. Clearly, our data would 
approach a $\delta$- function shape if plotted in terms of  $z$ defined in 
Eq.~(\ref{eq:pvsphi}).

\begin{figure}
{\centering \resizebox*{3.4in}{!}
{\includegraphics*{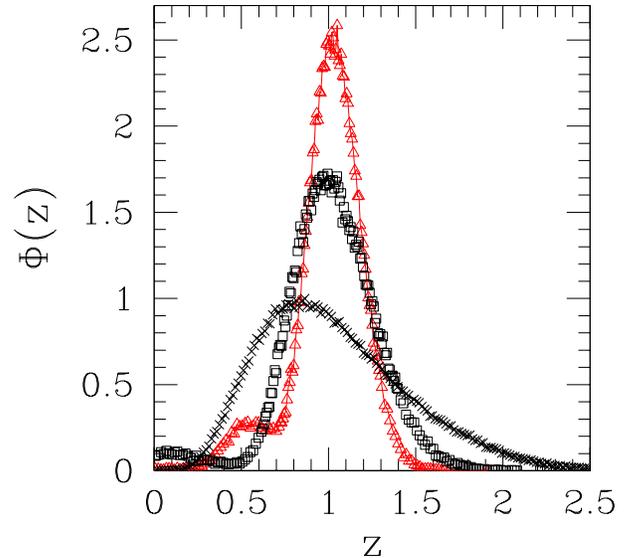}} \par}
\caption{(Color online) Scaled roughness distributions $\Phi (z)$ of time series, for 
$z$ of Eq.~(\protect{\ref{eq:pvsphi}}). $L=400$. Window width $W=10$ (triangles, $1.2 
\times 10^5$ samples), $2.5$ (squares, $1.2 \times 10^5$ samples), and 
$1.0$ (crosses, $5.7 \times 10^5$ samples).
}    
\label{fig:w10}
\end{figure}

Considering now narrower windows, and keeping the demagnetizing factor $\eta=0.005$, we 
show data for $W=10.0$, $2.5$, and
$1.0$ in  Fig.~\ref{fig:w10}, 
%%%%%%%%%%%%%%%%%%%%% 2nd ref's point # 6 %%%%%%%%%%%%%%%%%%%%%%%%%%%%%%%%%%%%%
where we have reverted to plotting our results in terms of 
the variable $z$ defined in Eq.~(\ref{eq:pvsphi}).
This is because it was noticed that, against diminishing $x$, the scaled roughness PDFs 
followed a trend away from the $\delta$-function shape which was the motivation for
using the variable $y$ of Eq.~(\ref{eq:y}).
%%%%%%%%%%%%%%%%%%%%%%%%%%%%%%%%%%%%%%%%%%%%%%%%%%%%%%%%%%%%%%%%%%%%%%%%%%%%%%%%%%%%%%%
In order to produce an accurate picture of
deviations from the Gaussian limit, we have generated a much larger number
of samples (${\cal O}(10^5)$) than for $W=100$. 
%%%%%%%%%%%%%%%% 1st ref's suggestion # 2 part II ########################################

\begin{figure}
{\centering \resizebox*{3.4in}{!}
{\includegraphics*{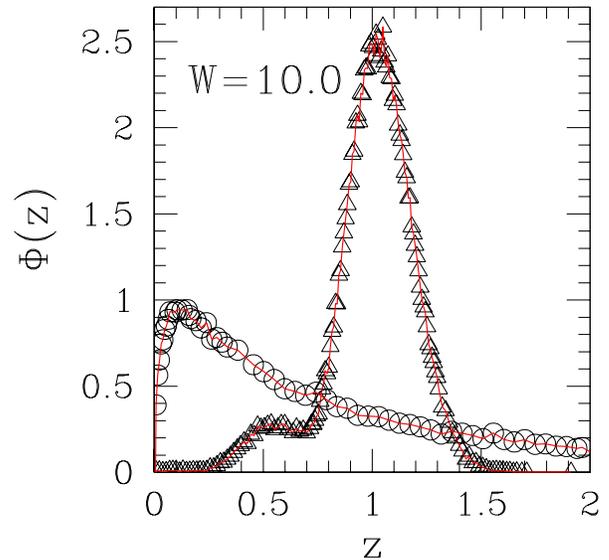}} \par}
\caption{(Color online) Scaled roughness distributions $\Phi (z)$ of time series, for 
$z$ of Eq.~(\protect{\ref{eq:pvsphi}}), with and without demagnetization. $L=400$, window 
width $W=10$. Triangles, 
$\eta=0.005$, $1.2 \times 10^5$ samples; circles, $\eta=0$, $2.1 \times 10^5$ samples.
}    
\label{fig:w10comp}
\end{figure}

Before analyzing the shapes exhibited in Fig.~\ref{fig:w10}, it is instructive to check  
how the demagnetization term influences the roughness PDFs
in the narrow-window limit. In Fig.~\ref{fig:w10comp} the scaled distributions 
for $W=10$ are shown, both with and without demagnetization. 
The shapes of PDFs are clearly rather distinct from each other, highlighting the 
relevance of demagnetization effects in this limit. For $\eta=0$ the distribution
peaks at $z \simeq 0.15$ and decays very slowly afterwards. As mentioned above in 
connection with the data of Fig.~\ref{fig:ltrt}~(b), this reflects the non-universal 
statistics of non-critical events which our calculational method for $\eta =0$
inevitably includes. The difference relative to that case is that for $W=10$,
each roughness sample is the  collation of only $\sim 19$ consecutive field sweeps. The 
corresponding results show that, in contrast to $W=100$, here one is outside the 
range of  applicability of the central limit theorem. 
%%%%%%%%%%%%%%%%%%%%%%%%%%%%%%%%%%%%%%%%%%%%%%%%%%%%%%%%%%%%%%%%%%%%%%%%%%%%%%%%%%%%%%%%

From now on we shall only deal with $\eta \neq 0$. 
Even though $W=10.0$ corresponds to $x \simeq 70$, it is 
clear from Fig.~\ref{fig:w10} that a secondary peak is evolving, i.e. a
significant distinction is emerging with respect to the simple
Gaussian picture found for larger $W$. Data for $W=5.0$ (not
shown) are virtually identical to those for $W=10.0$. While a
secondary peak still shows up for $W=2.5$, data for $W=1.0$ display only a
single maximum (however, these latter clearly differ from a pure
Gaussian).

\begin{figure}
{\centering \resizebox*{3.4in}{!}
{\includegraphics*{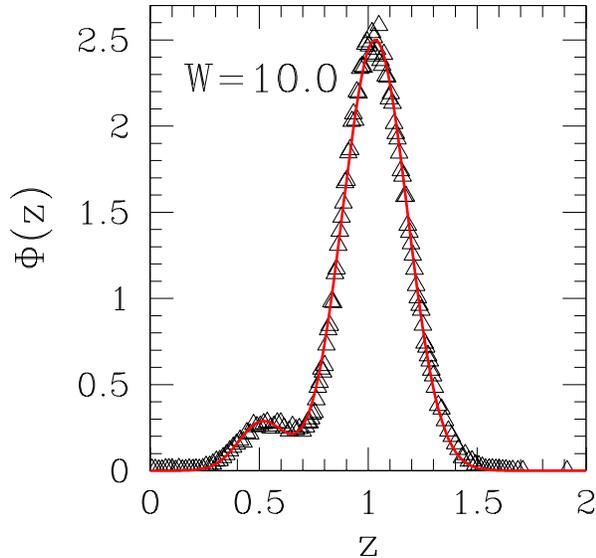}} \par}
\caption{(Color online) Scaled roughness distribution $\Phi (z)$ of time series, for $z$ 
of Eq.~(\protect{\ref{eq:pvsphi}}). $L=400$. Window width $W=10$. Triangles
are simulational data. Thick line is fit to Eq.~(\protect{\ref{eq:twog}}),
with $b=0.924(2)$, $a_1=1.03(1)$, $a_2=0.51(1)$. $\chi^2_{\rm \
d.o.f.} =1.5 \times 10^{-3}$~.
}    
\label{fig:w10fit}
\end{figure}

We then attempted to fit the data in Fig.~\ref{fig:w10} to analytical
forms. The $W=10.0$ results strongly suggest a double-Gaussian {\em
ansatz}, as:
\begin{equation}
\Phi(z)= b\,G_1(z) + (1-b)\,G_2(z)\ ,\\
\label{eq:twog}
\end{equation}
where $G_i$ is a Gaussian centered at $a_i$ with variance $\sigma_i^2$. 
As $W$ grows, one would expect $b \to 1$, $a_1 \to 1$ in
Eq.~(\ref{eq:twog}). Data for $W=10.0$ are well fitted by
$b=0.924(2)$, $a_1=1.03(1)$, $a_2=0.51(1)$, as seen in 
Fig.~\ref{fig:w10fit}.  The $\chi^2$
per degree of freedom ($\chi^2_{\rm \ d.o.f.}$) is $1.5 \times 10^{-3}$,
indicating that the form Eq.~(\ref{eq:twog}) indeed provides a
satisfactory description of simulational results in this case.

We have found that a similar fit, albeit of somewhat reduced quality
($\chi^2_{\rm \ d.o.f.} =3 \times 10^{-3}$, with  $b=0.955(5)$,
$a_1=1.02(1)$, $a_2=0.06(3)$) is feasible for the $W=2.5$
data as well. Turning to $W=1.0$, the double-Gaussian {\em ansatz}
worked surprisingly well, producing $\chi^2_{\rm \ d.o.f.} =6 \times
10^{-4}$, with  $b=0.53(5)$, $a_1=1.24(4)$, $a_2=0.77(1)$
(i.e. the two curves are roughly symmetric about $z=1$, with
approximately equal weights).  

Given that a double-peak structure is far from obvious for the $W=1.0$
data, alternative forms must be considered which might also provide a 
suitable fit to data in this limit. We investigated the family of
roughness PDFs for $1/f^\alpha$ noise~\cite{adgr02,rkdvw03}, keeping in
mind that window boundary conditions (WBC) are the appropriate ones in
this case~\cite{adgr01,adgr02,rkdvw03,mrkr04,rdist}. Such PDFs are
usually available in closed form~\cite{mrkr04}. However, close to
$\alpha=1$ it is more time-efficient to evaluate PDFs numerically via the
usual procedure of first generating a very long sequence of Gaussian
white noise, Fourier-transforming that sequence, multiplying the
Fourier components by $f^{-\alpha/2}$ and then inverting the
Fourier transform~\cite{adgr01,adgr02}. The resulting sequence is pure
$1/f^\alpha$ noise, which is then chopped into windows for analysis of
the corresponding roughness PDF. 

\begin{figure}
{\centering \resizebox*{3.4in}{!}
{\includegraphics*{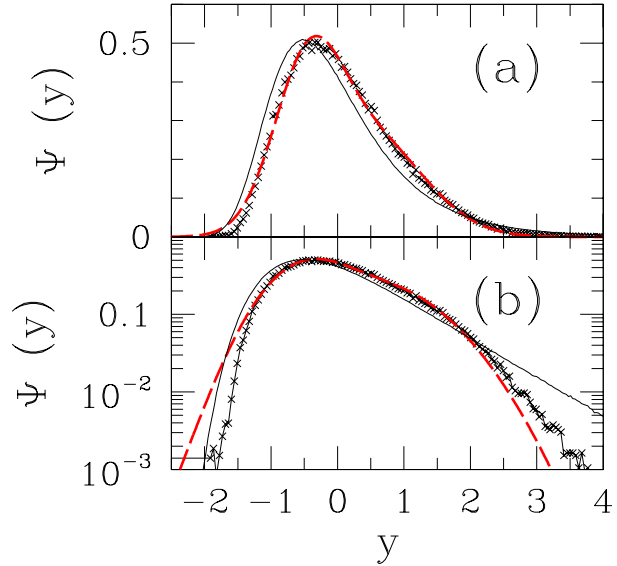}} \par}
\caption{(Color online) Crosses: scaled roughness distribution $\Psi (y)$ of time
series, for $y$ of  Eq.~(\protect{\ref{eq:y}}). $L=400$, window width
$W=1$. Dashed line is double-Gaussian fit to data 
(Eq.~(\protect{\ref{eq:twog}})). Full line is Fisher-Tippet-Gumbel
distribution with window boundary conditions.
Vertical axis is linear in (a), logarithmic in (b).
}    
\label{fig:w1fit}
\end{figure}

The best fit of the $1/f^\alpha$ family to our data was achieved for
$\alpha=1$, that is, the
Fisher-Tippet-Gumbel (FTG) statistics of extremes~\cite{adgr01}. Even so,
significant discrepancies remain. The overall picture is illustrated  in 
Fig.~\ref{fig:w1fit}, where we have switched again to the variable $y$ 
of Eq.~(\ref{eq:y}) because the FTG curve is better visualized in this 
way~\cite{adgr01,adgr02}.
One sees that, even though the double-Gaussian curve gives an excellent
fit in the central area of the plot where $\Psi(y) \gtrsim 0.1$, it fails
away from there, especially at the lower end. As to the FTG curve, while it follows 
the data closely, it never actually matches them.

\section{Discussion and Conclusions}
\label{conc}

The usual approach to the frequency domain in BN literature is via the study
of power spectra~\cite{dz05,ks00}. It has been found~\cite{us2} that, in the 
adiabatic limit of the interface model under consideration here, the power spectrum 
behaves approximately as $1/f^2$ within an intermediate range of frequencies.
One might construe this as indicating that the pure $1/f^2$ noise model of a Wiener 
process~\cite{adgr02,forwz94} applies in this case. However, the 
numerically-obtained full roughness  PDF, which contains much more information than 
a section of the power spectrum, tells a more nuanced story. Indeed, in general it 
does not follow a shape close to that of $1/f^\alpha$ curves, except for
narrow windows.
Even there, the closest fit within that family is for $\alpha \simeq 1$.

The question then arises of whether the generalized Gaussian approximation
underlying $1/f^\alpha$ noise models, in which the Fourier modes are considered 
as uncorrelated~\cite{adgr02}, is suitable for the description of BN time series.  
Our results, when considered in their evolution as window width varies,
appear more consistent with the idea that the similarity of our PDFs to
that of $1/f$ noise, found at the narrow-window limit, is fortuitous.
We recall that, even in studies of real-space interface roughness,
it is known that the independent-mode approximation gives rise to small 
but systematic discrepancies against experimental data, which can be traced
back to higher cumulants of the correlation functions~\cite{rkdvw03}.
Furthermore,  even more severe discrepancies have been found 
when boundary conditions other than periodic (e.g. window, as is the case here) are 
considered~\cite{rdist,rsk05}.

Turning now to the double-Gaussian picture, admittedly phenomenological 
in its inspiration, nonetheless it gives a description which is both numerically 
closer to actual data, and spans a broad range of window widths. 

The physical origins of the double-peak structure may be traced back to
the demagnetization term, and the consequent negative correlations
illustrated in Fig.~\ref{fig:tcorr}. A window of width $W$ contains at least
$W/\tau_M$ segments whose internal roughness profiles are uncorrelated to each 
other. On the other hand, within each such segment, negative correlations
are significant at least to some extent, thus preventing fluctuations 
from becoming very large. This latter effect gives rise to the secondary peaks  
at $y <0$, or equivalently, $z <1$. With $\tau_L \simeq 0.14$, 
$\tau_M \simeq 0.5$ for the $L=400$ systems which have been the focus of
our study, one has for $W=1$ that both inter- and intra-segment fluctuations
have similar weights, hence the $b \simeq 0.5$ result for the double-Gaussian
fit in that case. For $W/\tau_M \gg 1$ the dominant picture is one of 
many uncorrelated ``blobs'' of length $\sim \tau_M$, yielding the effective 
single-Gaussian limit observed.

\begin{figure}
{\centering \resizebox*{3.4in}{!}
{\includegraphics*{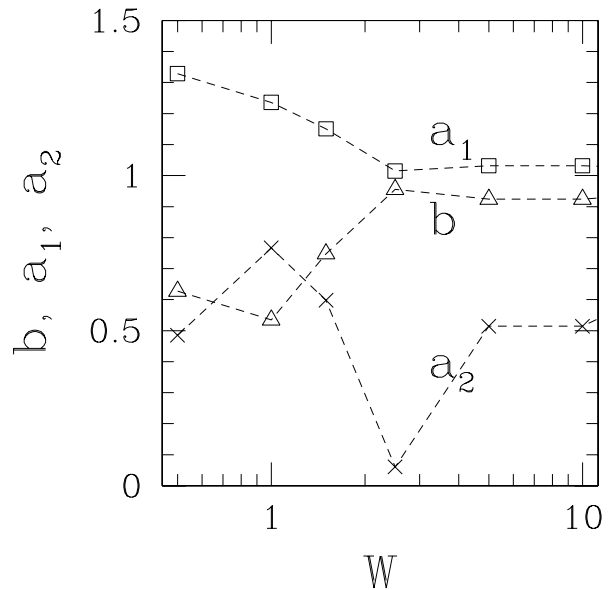}} \par}
\caption{Fitting parameters $b$ (triangles), $a_1$ (squares), and $a_2$
crosses) of double-Gaussian {\em ansatz}
of Eq.~(\protect{\ref{eq:twog}}), against window width $W$, for
$W=0.5$, $1.0$, $1.5$, $2.5$, $5.0$, $10$. $L=400$. 
}    
\label{fig:fitpar}
\end{figure}

%%%%%%%%%%%%%%%%%%%%%%%%%% 2nd ref's comment %%%%%%%%%%%%%%%%%%%%%%%%%%%%%
The double-Gaussian picture displays features which 
are not fully understood at present. Fig.~\ref{fig:fitpar} exhibits the
variation of parameters $b$, $a_1$, and $a_2$ of Eq.~(\ref{eq:twog})
against $W$ for not very large window widths (in addition to $W=1.0$,
$2.5$, $5.0$ and $10.0$ we ran simulations at $W=0.5$ and $1.5$).
While $b$ varies approximately as expected within this theoretical
framework (albeit with small non-monotonicities), 
and $a_1$ follows a rather monotonic trend, the behavior of $a_2$ is
intriguing, showing an apparent trend reversal. So far we have not able  
to provide an explanation for this.  

An alternative explanation for the observed behavior at $W \simeq 1$ may be 
proposed,  following a line similar to that advanced for the evolution of
$\eta=0$ data with  increasing $W$ (see Fig.~\ref{fig:w10comp} and the
respective discussion). In this scenario, the $W \simeq 1$ PDF shapes
would be non-universal (i.e. neither $1/f^\alpha$ nor
double-Gaussian). For larger $W \lesssim 10$  the central limit theorem
would  imply that, for the superposition of many (almost) 
decorrelated non-universal profiles, effective Gaussian structures 
should emerge. In  this view, the peak at $z <1$ would  again be ascribed
to segments within which negative 
correlations are  felt, with the peak at larger $z$ 
corresponding to inter-segment profiles.

Whatever the explanation of the behavior of roughness PDFs for $W \sim 1$, 
the extent of  window widths for which an effectively double-peaked structure 
shows up is considerably larger than, say, $\tau_L$. 
%%%%%%%%%%%%%%%%%%%%%%%%%%%%%%%%%%%%%%%%%%%%%%%%%%%%%%%%%%%%%%%%%%%%%%%%%%%%
Thus, a fairly 
straightforward way to detect the presence of demagnetization effects in 
experimental setups would be via the analysis of roughness PDFs of the induced
signal $V(t)$. Considering e.g. the conditions for the Perminvar samples described 
in  Ref.~\onlinecite{umm95}, where the average spacing between peaks is 13 msec
and $\tau_M \simeq 200$ msec, analysis of windows of width $\sim 2$ sec should
produce a well-defined double-peaked structure similar to that of 
Fig.~\ref{fig:w10fit}. 

\begin{acknowledgments}

This research  was partially supported by 
the Brazilian agencies CNPq (Grant No. 30.0003/2003-0), FAPERJ (Grant
No. E26--152.195/2002), FUJB-UFRJ and Instituto do Mil\^enio de
Nanoci\^encias--CNPq.
\end{acknowledgments}

\bibliography{biblio}  
\end{document}